\documentstyle[12pt,aaspp4]{article}

\received{}
\accepted{}
\journalid{}{}
\articleid{}{}

\slugcomment{Draft for ApJ}

\lefthead{Basri \& Mart\'\i n}
\righthead{Mass and Age of Faint $\alpha$~Per Objects}

\begin{document}

\title{The Mass and Age of Very Low Mass Members of the Open Cluster $\alpha$~Persei}

\author{Gibor Basri and Eduardo L. Mart\'\i n}
\affil{Astronomy Department, University of California,
    Berkeley, CA 94720}

\centerline{e-mail addresses: basri@soleil.berkeley.edu, 
 ege@popsicle.berkeley.edu} 

\begin{abstract} 

We present spectroscopic optical and photometric infrared observations of 12 faint candidate members of the young open cluster $\alpha$~Persei found by Prosser (\cite{prosser94}).  Keck HIRES echelle spectra provide radial and rotational velocity measurements for five objects, two of which are clearly nonmembers based on the radial velocities.  These kinematic nonmembers also do not fit well in the ($V-I$) vs ($I-J$) cluster sequence.  One additional faint object is likely a nonmember based on a low-resolution spectrum.

Using HIRES, we have searched for the Li\,{\sc i} resonance line.  Combining the absence/presence of lithium and photometry of the faint $\alpha$~Persei targets with confirmed membership constrains their ages and masses.  The lack of lithium in AP J0323+4853 implies its age is greater than about 65~Myr, which is older than the cluster classical upper main-sequence turnoff age of 50~Myr.  A similar age discrepancy is found in the Pleiades.  We detect lithium in the faintest of our program stars, AP270, which implies a mass for it just at the substellar mass limit, given our adopted age and its luminosity.  The membership of AP281 is in question because of its high radial velocity compared with the cluster mean.  On the other hand AP281 lies on the photometric cluster sequence, and has a very high rotation velocity and H$\alpha$ emission, indicating youth.  If a member, its lack of lithium would push the minimum age of the cluster to 75~Myr, in agreement with a very recent upper main-sequence determination. In that case, AP270 would not be a brown dwarf.

\end{abstract}


\section{Introduction}

Young open clusters have become a fertile hunting ground for very low-mass (VLM) star and brown dwarf (BD) surveys.  They offer the advantage that the VLM members are relatively bright and share the distance and metallicity derived from higher mass stars.  The successful identification of BDs in the Pleiades (Basri, Marcy \& Graham 1996, hereafter \cite{bmg}; Rebolo et al. \cite{rebolo96}) is a good example of how this advantage can be exploited.  The open cluster $\alpha$~Persei (Melotte 20) is one of the youngest open clusters within a radius of 200~pc from the Sun.  It is listed with an age of 50~Myr and a true distance modulus of m-M=6.36 in the catalogue of Lynga (\cite{lynga87}). Recently, the trigonometric parallax observations of the Hipparcos satellite have yielded a true distance modulus of 6.33$\pm$0.09 (Mermilliod et al. \cite{mermilliod97}), which is consistent with previous estimates.  Being younger than the Pleiades, and only a little more distant, $\alpha$~Per offers the opportunity of studying VLM objects at an earlier evolutionary stage.  

The ``lithium test'' for substellarity was first proposed and applied to a BD candidate found in $\alpha$~Persei (Rebolo, Mart\'\i n \& Magazz\`u \cite{rebolo92}).  Since then, it has been applied to BD suspects in the Pleiades and the field (Basri \cite{basri98a}), and has been used for constraining their ages and masses (\cite{bmg}).  In this work we  search for lithium in several of the faintest candidate members of $\alpha$~Per taken from Rebolo et al. (\cite{rebolo92}) and Prosser (\cite{prosser94}).  We find that the faintest of these objects (AP270 in Prosser's list) is indeed a `bona fide' cluster member and has a detectable lithium line.  We discuss the age and mass of this object using recent evolutionary models, and the methodology of ``lithium dating'' developed by \cite{bmg} (cf. Basri \cite{basri98b}). That work has shown that the Pleiades is substantially older than thought based on classical analysis of the upper main sequence turnoff (UMST) stars. Because lithium dating can provide a means of calibrating the convective overshoot in UMST stars (which affects their ages), it is important to confirm that this effect is generally seen in young clusters. By studying enough such clusters, the value of the overshoot parameter as a function of stellar mass could in principle be determined.

\section{Observations}

High-resolution spectroscopic observations of five of the faintest candidate members of $\alpha$~Persei from the list of Prosser (\cite{prosser94}) were obtained at the Keck~I telescope, using the HIRES echelle spectrometer (Vogt et al. \cite{vogt94}).  The instrumental configuration and data reduction procedure were the same as in \cite{bmg}.  With a dispersion of about 0.01 nm per (binned) pixel, the resolving power obtained was R$\sim$33000.  The targets observed are listed in Table~\ref{tab1}; the typical exposure time for the AP stars was about one hour.

Low-resolution spectroscopy of AP270, 275 \footnote{Original name AP J0323+4853 (Rebolo, Mart\'\i n \& Magazz\`u \cite{rebolo92}; Zapatero Osorio et al. \cite{zapatero96})} and AP276 was obtained on 1997 Dec 12 using the KAST double-arm spectrograph at the Shane 3-m telescope of the Lick observatory.  We removed the dichroic and employed the red arm only with grating number 6 (300~line~mm$^{-1}$).  The objects observed with KAST are listed in Table~\ref{tab2}.  The spectral resolution (FWHM) and wavelength coverage provided by this instrumental setup was 0.88 nm and 472.6--1023.2~nm, respectively.  Standard reduction and calibration procedures were accomplished with IRAF routines.  We corrected for the instrumental response using the flux standard HD~84937 (Stone \cite{stone96}).

Near-infrared observations of 12 candidate members from Prosser (\cite{prosser94}) were carried out on 1997 Sept 10 and 12 at the Lick 1~m telescope with the LIRC~2 camera (Gilmore, Rank \& Temi \cite{gilmore94}).  The detector is a NICMOS-3 256$^2$ pixel array with a field of view of 7.2 arcmin$^2$.  We used the standard $J$ and $K'$ filter set.  Total exposure times ranged from 60~s to 300~s in each filter.  We dithered the telescope making a square of 30 arcsec on the side.  The IRAF DAOPHOT package  was applied to reduce the frames and extract the photometry.  Instrumental aperture magnitudes were corrected for atmospheric extinction and transformed into the UKIRT system (Leggett \cite{leggett92}) using observations of the stars in the field of the Pleiades BD Calar~3 as standards (Zapatero Osorio, Mart\'\i n \& Rebolo \cite{osorio97}).  We found that our $K'$ photometry was useless because the background was too high and variable.  Thus, we give only the $J$-band photometry in Table~\ref{tab3}.  The ~1~$\sigma$ error is $<$0.1 mag.

\subsection{Spectroscopic analysis}

We measured heliocentric radial velocities for the five $\alpha$~Persei candidate members observed with HIRES by cross-correlating the spectra of three echelle orders with our spectrum of the M6V template Gl~406 obtained during the same run (on Dec. 7).  The spectral orders chosen cover the wavelength ranges  786.4--797.5~nm, 804.2--815.6~nm, and 822.9--834.6~nm which contain hundreds of sharp molecular lines (mainly TiO and VO). They are not contaminated by strong telluric lines.  The errors were estimated from the dispersion among radial velocities obtained from the different echelle orders, and are always $<$1 pixel. We adopt a radial velocity for Gl~406 of $v_{\rm rad}$=19.2$\pm$0.15 km s$^{-1}$ from X. Delfosse (personal communication). It supersedes his published value with a new determination based on four more observations and a better template. We also derived our own independent absolute radial velocity for Gl 406 using several atomic lines, obtaining 18.4$\pm$1 km s$^{-1}$. Our radial velocity determinations for the program stars are given in Table~\ref{tab1}. 

Our HIRES spectra are also useful for measuring the rotational broadening ({\it v}~sin{\it i}).  We used the same echelle orders as for the radial velocity measurements described above.  For calibration we convolved the spectrum of Gl~406 ({\it v}~sin{\it i}$<$2.9~km~s$^{-1}$ ; Delfosse et al. \cite{delfosse98})  with successive values of {\it v}~sin{\it i} in intervals of 10~km~s$^{-1}$, and cross-correlated these convolved spectra with the original spectrum.  The resulting correlation functions were compared with each of the correlation functions between a program star and the spectrum of Gl~406 to find the best the rotational velocity. In particular, a gaussian was fit to the core of each correlation function, and the gaussian width used to make the comparison. This procedure is similar to that employed by Basri \& Marcy (\cite{basri95}).  The {\it v}~sin{\it i} values obtained for the program stars are given in Table~\ref{tab1}. The errors reflect the dispersion in values obtained from the different orders, while the value itself is found by first averaging all the correlation functions for a given star.

The HIRES spectra were also used for measuring the strength of selected atomic lines.  Several interesting optical lines present in late M-type stars were included in our spectral range.  We focus on three features:  H$\alpha$, Li\,{\sc i} at 670.8 nm, and K\,{\sc i} at 769.9 nm, because they are indicators of chromospheric activity, nuclear burning in the interior, and surface gravity, respectively.  All our program stars showed H$\alpha$ in emission with single-peaked profiles.  The equivalent widths are given in Table~\ref{tab1}, along with upper limits for the equivalent width of the Li\,{\sc i} in the stars where we do not detect it. These were measured by making estimates of the pseudocontinuum by eye and taking a reasonable range of possibilities to estimate the errors. The spectra are noisy, and many features are not real, so the error estimate may be liberal. Only in AP270 is there a detection of lithium, as shown in Figure~\ref{fig1}. The apparent absorption features on either side of the line are not real, and its asymmetry should not be taken seriously. We note that our upper limits to the Li\,{\sc i} equivalent width are substantially lower than those obtained by Zapatero Osorio et al. (\cite{zapatero96}) from lower-resolution spectra of AP275 and AP279.  In Figure~\ref{fig2} we present the K\,{\sc i} region. Much of the blue wing of the K\,{\sc i} feature lies outside the recorded spectral domain. Nevertheless, we can clearly see that the K\,{\sc i} line in AP270 is narrower than in the M6 dwarf Gl406.  This difference is due to the lower gravity in the young $\alpha$~Per object compared to the older main sequence star.

We used our KAST spectra to measure pseudocontinuum ratios that are known to be good spectral type indicators for M-type stars (Mart\'\i n et al. \cite{martin96}).  We used the published values of the PC2, PC3, and PC4 indices for field dwarfs of known spectral type.  The PC1 and PC5 indices are within our spectral range but our spectra are too noisy at those wavelengths.  The mean spectral subclass and 1~$\sigma$ dispersion obtained from the PC2, PC3 and PC4 indices are shown in Table~\ref{tab2}.  For AP275 our spectral type is consistent with that of Zapatero Osorio et al. (\cite{zapatero96}).

\section{Discussion}

\subsection{Cluster membership}

According to Prosser (\cite{prosser92}, \cite{prosser94}), the mean heliocentric radial velocity of $\alpha$~Persei members is -2~km s$^{-1}$.  The radial velocities of AP270 and AP275 are close to the cluster mean, supporting membership.  On the other hand, AP273 and AP279 have much too high radial velocities, indicating nonmembership.  No radial velocity change larger than the measurement uncertainty was detected between the two spectra of AP273 (or those of AP270) taken on different nights.  Thus, it is unlikely that the high radial velocity of AP273 is due to a close companion.  The radial velocity of AP281 is higher than any of the high confidence $\alpha$~Per members.  Nevertheless, we consider this star a possible member because of its very high rotation velocity and good agreement with the cluster photometric sequence (see below).  It is conceivable that the radial velocity is variable due to an unresolved companion.  We do not find any trace of the companion in our spectrum so it would have to be much fainter than the primary (possibly a BD).  More high resolution spectra are needed in order to test this hypothesis.

A color-color diagram displaying the locus of kinematically young disk field dwarfs (Leggett \cite{leggett92}) is presented in Figure~\ref{fig3} together with the AP stars that we have measured at the Lick 1-m telescope.  We have dereddened the AP stars by the mean color excess of E(V-I)$_K$=0.16 (Prosser \cite{prosser94}).  Most of the AP stars are located around the young disk star locus.  It is interesting to note that the two most deviant stars are AP273 and AP279 which are nonmembers according to their radial velocities.  These examples show that color-color diagrams can be provide a very useful diagnostic of cluster membership, particularly in a cluster of low galactic latitude like $\alpha$~Per (b=-5.9$^\circ$) where there can be a significant amount of contamination from galactic disk stars.

We deem AP276 to be a nonmember of $\alpha$~Per because it has too early a spectral type for its (V-I) color.  Our final judgments concerning the membership of the AP stars studied in this paper are provided in Table~\ref{tab3}.  Stars for which we believe there is a high confidence of membership are flagged with `Y, while those with less confidence are flagged with `Y:', usually due to lack of kinematic information.  AP281 is kinematically a nonmember unless its radial velocity is variable.

\subsection{Activity and rotation}

Our results further confirm that $\alpha$~Persei VLM members have H$\alpha$ in emission as expected for their young age. This diagnostic in young clusters has been discussed by eg. Prosser, Stauffer \& Kraft \cite{prosser91}, and for $\alpha$~Per in particular by Prosser \cite{prosser92}. However, we note that the nonmembers AP273 and AP279 also have H$\alpha$ in emission.  Hence, H$\alpha$ does not seem to be a powerful criterion for cluster membership.  Zapatero Osorio et al. (\cite{zapatero96}) found that the maximum H$\alpha$ emission equivalent widths among $\alpha$~Persei stars occur around spectral type M4, and does not rise or even declines toward cooler members.  Our H$\alpha$ measurements are consistent with this.  A similar behavior has also been observed among VLM members of the Pleiades cluster (Hodgkin, Jameson \& Steele \cite{hodgkin95}; Stauffer, Liebert \& Giampapa \cite{stauffer95}).  The decrease in H$\alpha$ emission could be due to a reduction of the efficiency of the turbulent dynamo in generating chromospheric activity.  In field stars the decline of H$\alpha$ emission has also been observed, although at cooler temperatures (spectral type M7 or later; Basri et al. \cite{basri96}).  The difference in the spectral type at which H$\alpha$ emission becomes very weak between the young open clusters and the field VLM stars indicates that this phenomenon may be time-dependent.

The rotation velocities of the faintest $\alpha$~Per members observed by us are moderately high, which is not surprising because the stars are young.  The rotation velocity of AP275 is lower than the equatorial rotation velocity derived from the photometric period (Mart\'\i n \& Zapatero Osorio \cite{martin97}), indicating that the star is viewed at a moderate inclination.  What is surprising is that the radial velocity nonmembers AP279 and especially AP281 have rather high {\it v}~sin{\it i}. Rotational velocities $\ge$20~km~s$^{-1}$ are unusual among M5--M6 dwarfs in the solar neighborhood (Delfosse et al.  \cite{delfosse98}).  

Since AP279 and AP281 were discovered in a photometric survey, which is unbiased with respect to rotation or chromospheric activity, there is no reason that they should be young if they do not belong to the cluster.  As noted above, AP281 might be cluster member if it is a spectroscopic binary.  But this possibility is unlikely for AP279 because it fails to match the cluster sequence in the color-color plot (Figure~\ref{fig3}).  We do not have a firm explanation for why these stars are rotating so fast if not members of $\alpha$~Per.  The Hyades ($\sim$600 Myr) contains some M5 members with {\it v}sin{\it i} in the range 20--40~km~s$^{-1}$ (Jones, Fischer \& Stauffer \cite{jones96}).  Even higher velocities appear to be the norm at the age of the Pleiades (Oppenheimer et al.  \cite{oppenh97}).  The latter paper discusses stars which satisfy almost all cluster membership criteria, yet are very unlikely to be members.  We can speculate (as did Oppenheimer et al. \cite{oppenh97}) that there is an enhancement in the number of relatively young field stars in the general direction of these northern winter clusters.

\subsection{The age of the cluster}

\subsubsection{Isochrone Fitting}

The location of our program stars in a color-magnitude diagram is shown in  Figure~\ref{fig4}. We have also plotted the positions of Pleiades stars  from the proper motion study of Hambly, Hawkins \& Jameson (\cite{hambly93}) with IR photometry provided by Steele, Jameson \& Hambly  (\cite{steele93}). Also shown are Pleiades BDs with lithium detections (Rebolo et al. \cite{rebolo96}; Mart\'\i n et al. \cite{martin98}) including IR photometry from Zapatero Osorio et al. (\cite{osorio97}).  If we disregard the kinematic nonmembers AP273 and AP279, the rest of the AP objects define  a relatively narrow sequence in this diagram, including the uncertain member AP281.  On the other hand, the Pleiades  stars occupy a much wider region. The $I$-band photometry for  many of these stars comes from photographic plates and could be  inaccurate (Steele \& Jameson \cite{steele95}).  It is also possible  that there are contaminating field stars in the Pleiades sample, even  though it has been selected from a proper motion study. For instance,  Mart\'\i n et al. (\cite{martin96}) recently found that HHJ~7 is a nonmember on the basis of low-resolution spectroscopy, and several  other HHJ stars are suspected nonmembers because they lie below the  main-sequence. Furthermore, Steele \& Jameson (\cite{steele95}) have shown that part of the spread is due to unresolved binaries.

The $\alpha$~Persei stars overlap with the brighter half of their Pleiades counterparts, suggesting that it is on the average a younger  sample. However, the scatter seen in the Pleiades prevents a more  detailed comparison. Theoretical isochrones for 30, 50, 70 and 100~Myr  (Baraffe et al. \cite{baraffe98}) computed using NextGen model atmospheres (Allard et al. \cite{allard98}) are superimposed on the data in Figure~\ref{fig4}. The isochrones are parallel to the general observed sequence. All the $\alpha$~Per members  lie above the 70~Myr isochrone, and most are even above the 50~Myr one.  However, we do not think that $\alpha$~Per stars are younger than 50~Myr  because it would not be consistent with any other age estimates, as  discussed below. We note that there could be problems with the  isochrones of  Baraffe et al. (\cite{baraffe98}) mainly due to  the water absorption line list which affects the IR colors (Allard et al. \cite{allard94}, \cite{allard98}). 

The dispersion of the data points in Figure~\ref{fig4} might suggest that there could be a spread in ages  among $\alpha$~Persei members of $>$20~Myr. However, this cannot be  taken at face value because there are several factors producing  uncertainties in estimating the ages of individual objects from a  color-magnitude or an H-R diagram. In addition to the sources of  dispersion already cited above for the Pleiades (photometric errors and contamination from field stars), there could easily be some unresolved binaries.  It is also the case that these  young stars are intrinsically variable.  Mart\'\i n \& Zapatero Osorio (\cite{martin97}) have shown that the VLM  $\alpha$~Per members can have brightness changes of order 0.1--0.2 mag.  in the $R$ and $I$ filters.  For AP275 they found average magnitudes  of $I_c=17.13$ in their 1994 data, and $I_c=17.03$ in their 1995 data,  while Prosser (\cite{prosser94}) obtained $I_c=17.22$ in his 1991 data. We have adopted  the value of $I_c=17.13$ because it is intermediate between the other two and  has a lower formal error. 

\subsubsection{Lithium Dating}

A more precise method of dating VLM cluster members is by `lithium dating' (\cite{bmg}; Bildsten et al. \cite{bildsten97}; Basri \cite{basri98b}). This exploits the fact that the luminosity at which lithium is efficiently  depleted in fully convective objects is a steep function of time and fairly model independent. Using analytical calculations,  Ushomirsky et al. (\cite{usho98}) have obtained an equation which  gives a lower limit on the age of a non-degenerate  fully convective star that has depleted lithium, solely as a function  of its luminosity and metallicity. They applied their results to  the  $\alpha$~Persei stars with lithium nondetections from  Zapatero Osorio et al. (\cite{zapatero96}) and inferred a minimum age of 61$\pm$7~Myr for the cluster (if coeval) from AP279. Unfortunately,  we have found AP279 to have a radial velocity inconsistent with cluster  membership. The next faintest $\alpha$~Per star with a radial velocity fully consistent with membership and a lithium nondetection is AP275.  Using Figure~4 of Ushomirsky et al. (\cite{usho98}), and the range  in luminosities estimated by Zapatero Osorio et al. (\cite{zapatero96}),  we find a minimum age of 58$\pm$8~Myr. This age estimate is also  quite insensitive to the actual lithium abundance, as demonstrated  in equation 40 of Ushomirsky et al. (\cite{usho98}). It does depend on the effective temperature calibration for these pre-main sequence objects, which is still subject to some uncertainty, and on the assumption of evolution at constant effective temperature needed for the analytic analysis.

Here we perform the analysis with detailed models and current observations.
We have primarily used the evolutionary models of Baraffe et al.
(\cite{baraffe98}), kindly supplied to us in advance of publication by I.
Baraffe (cf.  Chabrier \& Baraffe \cite{chabrier97}).  It is reasonable to use
these models because they provide a good fit to the photometric cluster
sequence of $\alpha$~Per members.  A further advantage is that the authors
directly provide the theoretical absolute magnitudes in several passbands
because they use model atmospheres which provide reasonable good fits to the
observed spectra of VLM stars and BDs (Allard et al.  \cite{allard98}).  In
Figure~\ref{fig5} we present the lithium depletion predictions in the $I_c$ and $J$ filters.  For the latter filter we have transformed the calculations of Baraffe et al. (\cite{baraffe98}) from the CIT system to the UKIRT system
using the relationship provided by Leggett (\cite{leggett92}) derive from US Naval Observatory data.  We adjust the observed $I_c$ and $I_c-J$ magnitudes for reddening with corrections of -0.14 and -0.16 magnitudes respectively.  

The following values of true distance modulus for $\alpha$~Per were found in
the literature:  6.36 (Lynga \cite{lynga87}), 6.4 (Prosser \cite{prosser94}),
6.24 (Dzervitis et al.  \cite{dzervitis94}) and 6.33 (Mermilliod et al.
\cite{mermilliod97}).  These have been corrected with the values of reddening
adopted by each author.  The mean value is m-M=6.33, which coincides with the
Hipparcos-based result of Mermilliod et al.  (\cite{mermilliod97}). The
Hipparcos distance scale has been called into question for some young clusters
(Soderblom et al.  \cite{soderblom98}), but the main sequence found for
$\alpha$~Per is reasonable.  We adopt a true distance modulus of 6.33 for this
paper, with a presumed error of 0.1 magnitudes.

Bolometric luminosities of the stars can be derived using two different methods. One is to use the empirical bolometric corrections derived from main sequence stars with the same $V-I$ colors, obtained from Monet et al.  (\cite{monet92}). The other (which we adopt) is to use the bolometric correction for $I$ calculated by Baraffe et al. (\cite{baraffe98}). This has the advantage of working directly from $I$, without needing to assume that the colors of the stars are the same as main sequence stars. It has the disadvantage that we must rely on the model spectra. These are expected to be less trustworthy in the $J$ band due to incomplete water and other molecular line lists (Allard et al. \cite{allard94}, \cite{allard98}). This could be the source of the offset of the isochrones in Fig.~\ref{fig4} discussed earlier. Moving the isochrones about 0.1 mag redward in $I_c-J$ would make the age inferred from them in closer agreement to the lithium age, and at the same time make the lithium age inferred from $J$ (Fig. 5b) agree better with that from $I_c$ (Fig. 5a) or bolometric luminosity (Fig.~\ref{fig6}). 

The combination of the absolute $I_c$ and $J$ magnitudes and luminosities of AP270, AP275, and AP281 with the theoretical predictions for lithium depletion yields the following results: 

\begin{itemize}

{\item For an age of 50~Myr, which is the canonical cluster age  based on the UMST stars (Mermilliod  \cite{mermilliod81}), AP270 is expected to have preserved more than  90\% of its initial lithium content, consistent with our lithium detection in it. However,  AP275 is expected to  have preserved 50\% (Fig. 5a -- $I$ band) or 10\% (Fig. 5b -- $J$ band)  of its initial lithium. This is not  consistent with our upper limit to the lithium equivalent width, which  implies a lithium depletion of more than 2 orders of magnitude according to the  calculations of  Pavlenko et al. (\cite{pavlenko95}) and Pavlenko (\cite{pavlenko97}).  The initial lithium abundance of  $\alpha$~Per  members is the standard ISM value in the stars hotter than about  5500~K (Randich et al. \cite{randich98}). If AP281 is a member then its lack of lithium is also inconsistent with an age of 50~Myr. }

{\item For an age of 65~Myr, AP270 is expected to have preserved about 90\% (Fig. 5a) or 50\% (Fig. 5b)  of its initial lithium content, and AP275 should have preserved  less than 1\%. Both expectations are consistent with our lithium  observations. The age at which AP275 depletes lithium is a little older (5~Myr) than the estimate based on the analytical calculations of Ushomirsky et al. (\cite{usho98}). If, however, AP281 is a member then it would be expected to still have 50\%(10\%) of its lithium, and we should still see a strong lithium line in it. Thus, the minimum cluster age is 65~Myr based on AP275, but must be older if AP281 is a member. }

{\item For an age of 75~Myr, AP270 is still expected to preserve 50\%  (Fig. 5a) or 10\% (Fig. 5b) of  its initial lithium abundance. Our  measured lithium equivalent width implies a surface  lithium abundance of more than 10\% of the original according to the `pseudoequivalent' width calculations  of Pavlenko (\cite{pavlenko97}). Hence, AP270 is still reasonably consistent with such an age. Both AP275 and AP281 would have depleted lithium at this age, but this is about how long it takes a star with the brightness of AP281 to do it. Thus, the minimum cluster age if AP281 is a member is roughly 75~Myr, or about 50\% older than the age given by the classical UMST stars. These conclusions are all fully consistent with what is found if luminosities instead of $I_c,J$ magnitudes are used, as can be seen in Fig. \ref{fig6}. }

\end{itemize}

While the lithium nondetection in AP275 (or AP281) sets a minimum age for the cluster, the lithium  detection in AP270 sets a maximum age. This is illustrated in  Figure~\ref{fig6}, in which the 1\% lithium line can be taken as essentially the dividing line between spectroscopic detection or nondetection of lithium. The choice of 1\% does not matter much, since the entire process of lithium depletion takes place in a few Myr. The point where this line is intersected by the luminosity of a given star defines the minimum/maximum age of the star depending on whether lithium is a nondetection/detection. For AP270 the lines intersect a little below 90 Myr, so that is the maximum allowed age of the cluster based on these data. It would be surprising if it were that old, of course. Error bars on the luminosity based on the uncertain distance modulus are also shown in Fig.~\ref{fig6}. These translate into age uncertainties by affecting the point at which the luminosity intersects the lithium depletion line. It can be seen from the Figure that the uncertainty in age from this source is about $\pm5$Myr. 

We have tried to assess what the uncertainties due to the models are. In addition to using the Ushomirsky et al. (\cite{usho98}) formula and the Baraffe et al. (\cite{baraffe98}) models, we have been kindly provided models by A. Burrows (private comm., cf. Burrows et al. \cite{burrows97}). We compared the luminosity evolution of objects of different mass between these and those from  Baraffe, and the ages at which lithium disappears in the two models. The Burrows calculations tend to be a little brighter and hotter for a given mass and age; apparently their atmospheric treatment leaks flux a little more slowly than the Baraffe treatment. This leads to later times for lithium depletion in the Burrows models, by about 10~Myr in the relevant range of mass and age. We do not wish to speculate on which set is more ``correct'', preferring to note that for our purposes the Baraffe models yield more conservative conclusions.
 
The discussion above leads us to adopt an age for the $\alpha$~Persei cluster of 65~Myr as consistent with our lithium dating. The age is unlikely to be lower than 60~Myr, and could be as high as 75~Myr or more if AP281 is a member (or if other stars of similar brightness and better pedigree are found not to have lithium). In the Pleiades cluster lithium dating has led to an age estimate for the members around the substellar limit of 115$\pm$10~Myr (\cite{bmg};  Mart\'\i n et al. \cite{martin98}), which is about 1.5 times older than the canonical  age (Mermilliod \cite{mermilliod81}). This is roughly the same effect that we are finding for $\alpha$~Per. As discussed by \cite{bmg} and Basri (\cite{basri98b}),  this discrepancy can be solved with convective core overshooting in  evolutionary models of UMST stars. Meynet, Mermilliod \& Maeder  (\cite{meynet93}) computed models with moderate overshooting and obtained  older ages for the galactic open clusters. In particular, they found  100~Myr for the Pleiades and 52~Myr for $\alpha$~Persei. These ages are  more in agreement with the ages inferred from lithium in VLM members than the previous ones obtained without overshooting. 

The lithium  results actually indicate that the cluster ages are somewhat older than the  Meynet et al. (\cite{meynet93}) scale, suggesting that stronger overshooting  is necessary in their models. On the other hand, the overshooting employed by Mazzei \& Pigatto (\cite{mazzei89}) was too strong because it yielded a Pleiades age of 150~Myr.  Very recently Ventura et al. (\cite{ventura98}) have redone the convective overshoot calculations using the ``full spectrum turbulence'' treatment of Canuto, Goldman \& Mazzitelli (\cite{cgm96}). With approximately the same amount of overshoot as Maeder et al. they find the same ages for the Pleiades and $\alpha$~Per as we do (presuming AP281 is indeed a member). While such good agreement may be fortuitous, it does indicate that the lithium dating technique is robust, and that the source of its disagreement with classical UMST ages almost certainly lies in the stellar evolution treatment of high mass stars. 

We conclude that the lithium observations  of VLM cluster members can provide the best way of empirically calibrating  the amount of convective overshooting in high-mass stars, as first suggested by \cite{bmg}.  Obviously we should continue trying to define the lithium boundary in open clusters of various ages to refine these conclusions, and in $\alpha$~Per itself it is important to locate a number of other stars near the substellar boundary and test them for lithium. 

\subsection{Is AP270 a brown dwarf?}

The mass of any VLM cluster member can be estimated from its luminosity, once the cluster age is established (Fig.~\ref{fig6}). With our adopted age of 65~Myr, the mass of AP275 is a little over 0.1~M$_\odot$, the mass of AP270 is right at the substellar limit of 0.075~M$_\odot$ (75 jupiters), and if AP281 is a member its mass would be around 0.087~M$_\odot$. If the cluster is yet older, then all 3 objects are stellar. On the other hand, if the age is any less than 65~Myr, then AP270 moves more comfortably into the substellar domain. The boundary of lithium reappearance is not expected to coincide with the substellar boundary in this cluster; at the bottom right corner of Fig.~\ref{fig6} the depletion line crosses the substellar boundary at about 120~Myr (the age of the Pleiades). Thus we would expect stars a little brighter than AP270 to also show lithium. Alternatively, if the cluster is 75~Myr old then AP270 is itself a star which shows lithium. In any case, AP270 is a benchmark object defining the reappearance of lithium in the $\alpha$~Persei open cluster. Brown dwarf candidates fainter  than AP270 have not been identified yet. They should be well within the capabilities of present CCD imaging systems on 2~m class telescopes, and we are conducting a program to find them. 

The mass estimated from the presence of lithium in AP270 allows us to test the predictions of  theoretical evolutionary models at the  substellar boundary. For an age  of 65~Myr and a mass of 0.075~M$_\odot$, the Baraffe models predict absolute magnitudes of  M$_V$=14.55, M$_I$=11.50, and M$_J$=9.37; which can be compared with  M$_V$=15.24, M$_I$=11.40, and M$_J$=9.25 for AP270 obtained  using a distance modulus of 6.33 and an interstellar reddening of A$_V$=0.3, E(V-I)=0.16.  The predicted magnitudes are only 0.1 mag faint in $I_c$ and $J$, but for $V$ the predicted magnitude is uncomfortably bright. The use of atmospheric opacities which include the effects of dust may help to reduce the discrepancy in $V$. Increasing the brightness of the models in $I$ and $J$ would help in Fig.~\ref{fig4} to reconcile the isochrone age with the lithium age. This would also help with the problem that the faintest 2 $\alpha$~Per members appear to be substellar in Fig.~\ref{fig4}, while we have argued above that they are not so low in mass. The increase needed is 0.3 magnitudes, however. The Burrows models are brighter, but we do not have the specific color predictions which would allow us to make the comparison with our photometry. 
   
A fine analysis of our high-resolution spectra using spectral synthesis  should be able to determine the surface lithium abundance of  AP270.  The same kind of analysis should provide good estimates of the  temperature and gravity of AP270, which would be very useful for testing  evolutionary models. It would also result in improved estimates of its age and mass. We are beginning a larger program of fine analysis for several of the currently known BDs, using models of Allard and Hauschildt including the effects of dust.

\acknowledgments

{\it Acknowledgments}: 
This research is based on data collected at the  W.~M. Keck Observatory, which is operated jointly by the University of  California and California Institute of Technology, and on the Shane 3-m and Nickel 1-m telescopes  at Lick Observatory run by the University of California.  GB acknowledges the support of NSF through grant AST96-18439. EM acknowledges support from the Fullbright-DGES program  of the Spanish ministry of Education. We gratefully thank Isabelle Baraffe and Adam Burrows for sending theoretical modeling results in advance of publication. 

%\clearpage

\clearpage

\begin{deluxetable}{lccccccc}
\footnotesize
\tablecaption{\label{tab1} HIRES data}

\tablewidth{0pt}
\tablehead{
\colhead{AP}  & 
\colhead{UT Date}  &
\colhead{RJD 2450+}  &
\colhead{S/N}  & 
\colhead{$v_{\rm rad}$}  &
\colhead{{\it v}~sin{\it i}}  &
\colhead{EW (Li\,{\sc i})}  &
\colhead{EW (H$\alpha$)}  \nl   
 &  &  &  & km s$^{-1}$ & km s$^{-1}$ & \AA & \AA \nl}

\startdata
270 & 2 Dec 1997 & 784.85& 7 & -2.0$\pm$1.8 & 24$\pm$4 & 0.62$\pm$0.06 & -5.3$\pm$0.5 \nl
270 & 7 Dec 1997 & 789.80& 9 & -0.4$\pm$1.5 & - & 0.70$\pm$0.08 & -4.5$\pm$0.5 \nl
273 & 3 Dec 1997 & 785.90& 2 & 26.2$\pm$2.1 & 7$\pm$3 & $<$0.33 & -3.8$\pm$0.6 \nl
273 & 7 Dec 1997 & 789.85& 4.5 & 25.3$\pm$0.8 & - & $<$0.27 & -5.9$\pm$0.3 \nl
275 & 1 Dec 1997 & 783.85& 4.5 & -2.0$\pm$1.3 & 33$\pm$6 & $<$0.25 & -8.1$\pm$0.3 \nl
279 & 2 Dec 1997 & 784.85& 10 & 33.0$\pm$1.8 & 20$\pm$6 & $<$0.26 & -2.2$\pm$0.4 \nl
281 & 7 Dec 1997 & 789.75& 12 &  9.4$\pm$1.2 & 51$\pm$5 & $<$0.14 & -3.5$\pm$0.3 \nl
\enddata
\tablenotetext{}{The approximate S/N is given for the order centered at 849 nm. It decreases to shorter wavelengths. It is hard to estimate precisely for cool stars like these because the molecular lines themselves look like noise.}
\end{deluxetable}

\begin{deluxetable}{lcccccc}
\footnotesize
\tablecaption{\label{tab2} KAST data}
\tablewidth{0pt}
\tablehead{
\colhead{AP}  & 
\colhead{S/N}  & 
\colhead{PC1}  &
\colhead{PC2}  &
\colhead{PC3}  &
\colhead{PC4}  &  
\colhead{SpT}  \nl   
                            }
\startdata
270 & 20 & 1.55 & 2.49 & 1.49 & 1.70 & M6.7$\pm$0.8 \nl
275 & 17 & 1.79 & 2.24 & 1.30 & 1.91 & M6.2$\pm$0.6 \nl
276 & 13 & 1.81 & 1.67 & 1.17 & 1.86 & M5.1$\pm$0.6 \nl
\enddata
\tablenotetext{}{S/N is for wavelengths around 810 nm.}
\end{deluxetable}

\begin{deluxetable}{lccccccc}
\footnotesize
\tablecaption{\label{tab3} Photometric data}
\tablewidth{0pt}
\tablehead{
\colhead{AP}  & 
\colhead{V}  &
\colhead{$(V-I_c)_0$}  & 
\colhead{$J_{uk}$}  &
\colhead{$(I_c-J)_0$} &
\colhead{M$_{I_c}$} &
\colhead{log(L$_{bol}$)} &
\colhead{Membership} \nl   
                            }
\startdata
268 & 20.50 & 3.43 & 14.98 & 1.77 & 10.44 & -2.35 & Y:\nl
270 & 21.87 & 3.84 & 15.56 & 2.15 & 11.40 & -2.66 & Y \nl
272 & 20.42 & 3.58 & 14.71 & 1.81 & 10.21 & -2.27 & Y:\nl
273 & 21.71 & 3.76 & 15.33 & 2.30 & 11.32 & -2.64 & N \nl
275 & 21.10 & 3.81 & 14.95 & 2.09 & 10.64 & -2.45 & Y \nl
276 & 21.60 & 3.69 &  -    &  -   & 11.28 & -2.63 & N \nl
279 & 21.10 & 3.66 & 15.73 & 1.39 & 10.81 & -2.48 & N \nl
281 & 21.41 & 3.70 & 15.45 & 1.94 & 11.08 & -2.56 & ? \nl
282 & 18.48 & 2.90 & 13.75 & 1.51 &  8.95 & -1.82 & Y:\nl
284 & 19.44 & 3.23 & 14.33 & 1.56 &  9.58 & -2.05 & Y:\nl
290 & 19.20 & 3.23 & 14.04 & 1.61 &  9.34 & -1.96 & Y:\nl
297 & 17.63 & 2.71 & 13.16 & 1.44 &  8.29 & -1.57 & Y:\nl
\enddata
\tablenotetext{}{The V photometry is from Prosser (1994). The $V-I$ color is also from Prosser, transformed from the Kron to the Cousins system using the cubic equation given in Leggett (1992). For AP275 (=AP J0323+4853) we adopted the average NOT I-band photometry provided by Mart\'\i n \& Zapatero Osorio (1997). The colors and M$_{I_c}$ values have been corrected for reddening. Luminosities are in solar units, estimated using M$_{I_c}$. }
\end{deluxetable}

\clearpage

\plotone{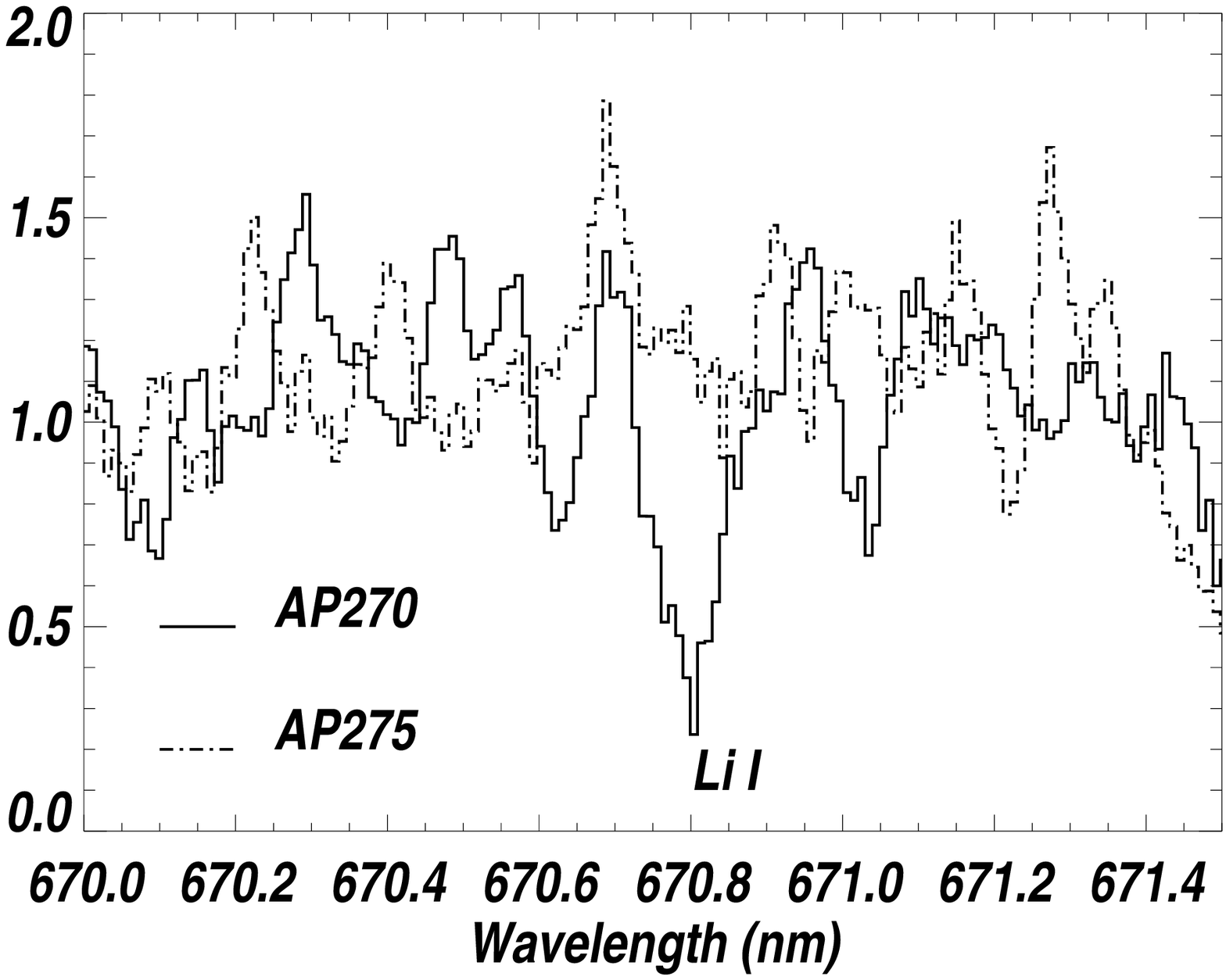}

\figcaption[li270.ps]{\label{fig1} HIRES spectra of AP270 (solid line) 
and AP275 (dashed line) around the Li \,{\sc i} 670.8 nm ~ resonance 
line. We have applied a boxcar smoothing of 5 pixels. } 

\plotone{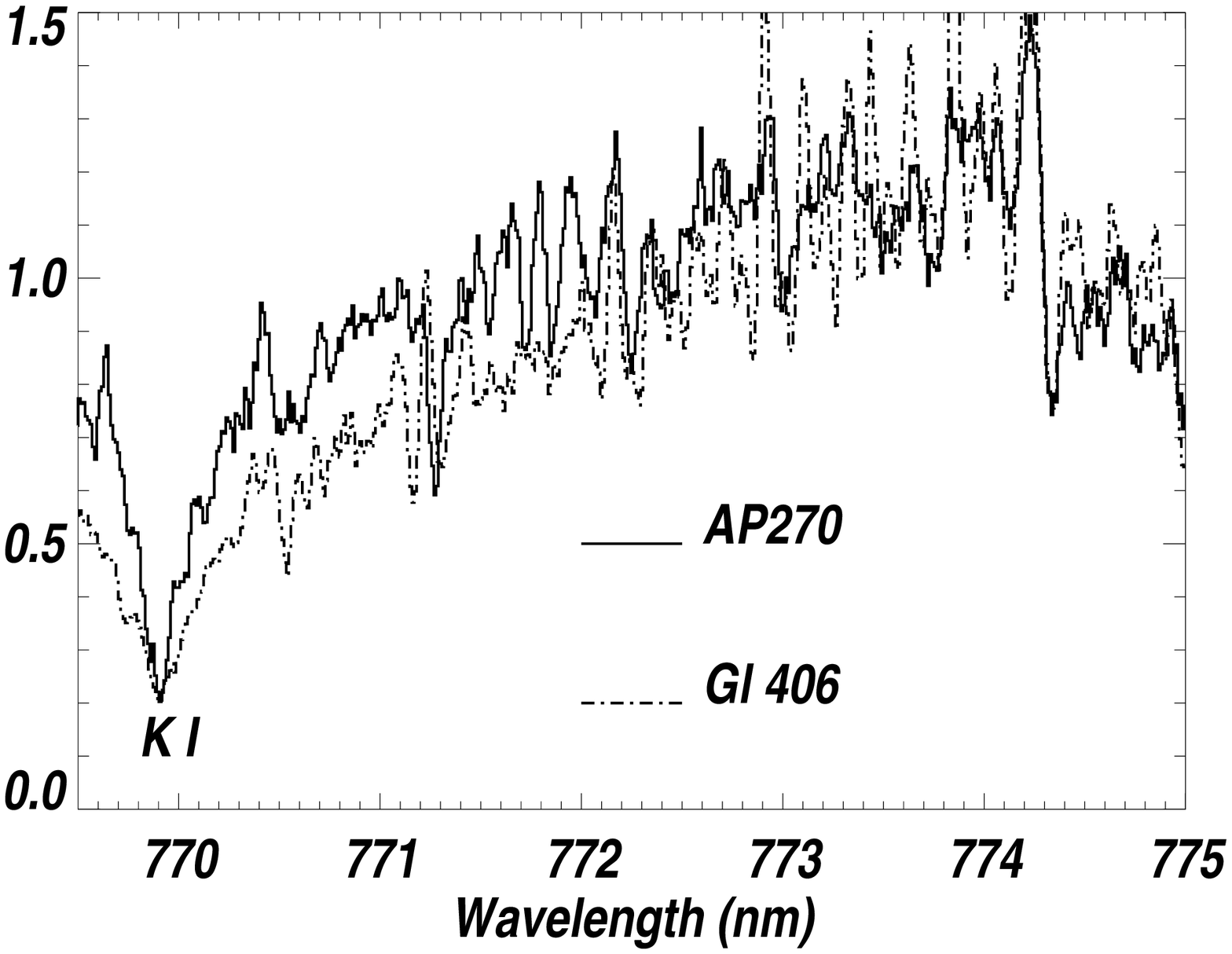}

\figcaption[k270.ps]{\label{fig2} HIRES spectra of AP270 (solid line) 
and Gl406 (dashed line) around the K \,{\sc i} 769.9 nm ~ resonance 
line. We have applied a boxcar smoothing of 3 pixels to the spectrum 
of AP270. We have convolved the spectrum of Gl406 with a rotational broadening of 17.5 km s$^{-1}$, and we have normalized it to the same counts as the spectrum of AP270 in the the spectral region 773--780~nm .} 

\plotone{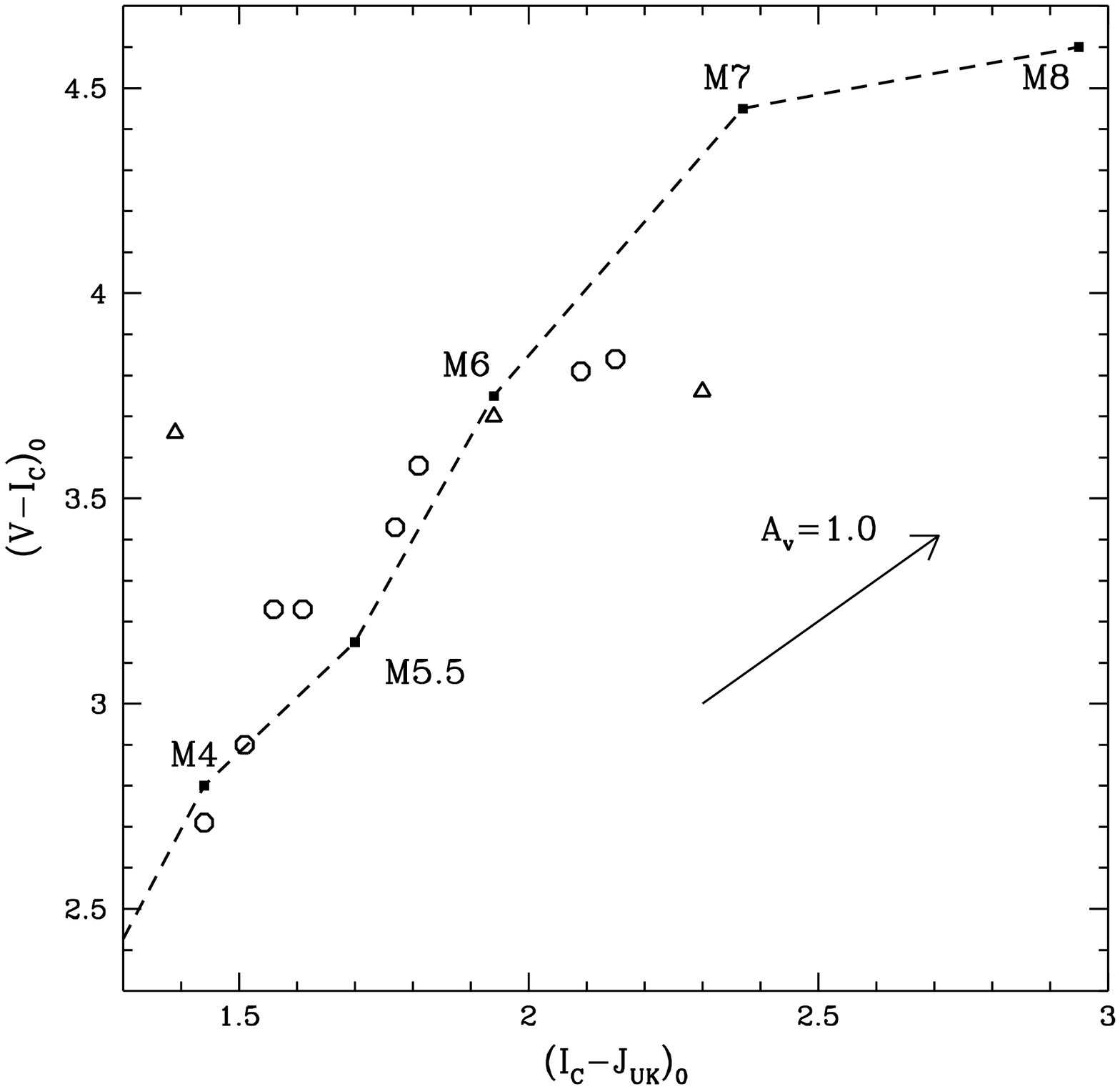}

\figcaption[color.eps]{\label{fig3} The ($V-I_c$) vs ($I_c-J$) sequence in 
$\alpha$~Persei. Open symbols are the candidate members from Prosser (1994) with J-band photometry obtained by us. The open triangles are the  $\alpha$~Persei radial velocity nonmembers (AP273, AP279 and AP281). 
Filled squares represent the sequence for young disk field dwarfs provided by Leggett (1992). Their spectral types are labeled.  The arrow indicates the color effect of 1 mag. of visual reddening. A constant reddening of E(V-I)=0.16 has been applied to the $\alpha$~Persei stars.}

\plotone{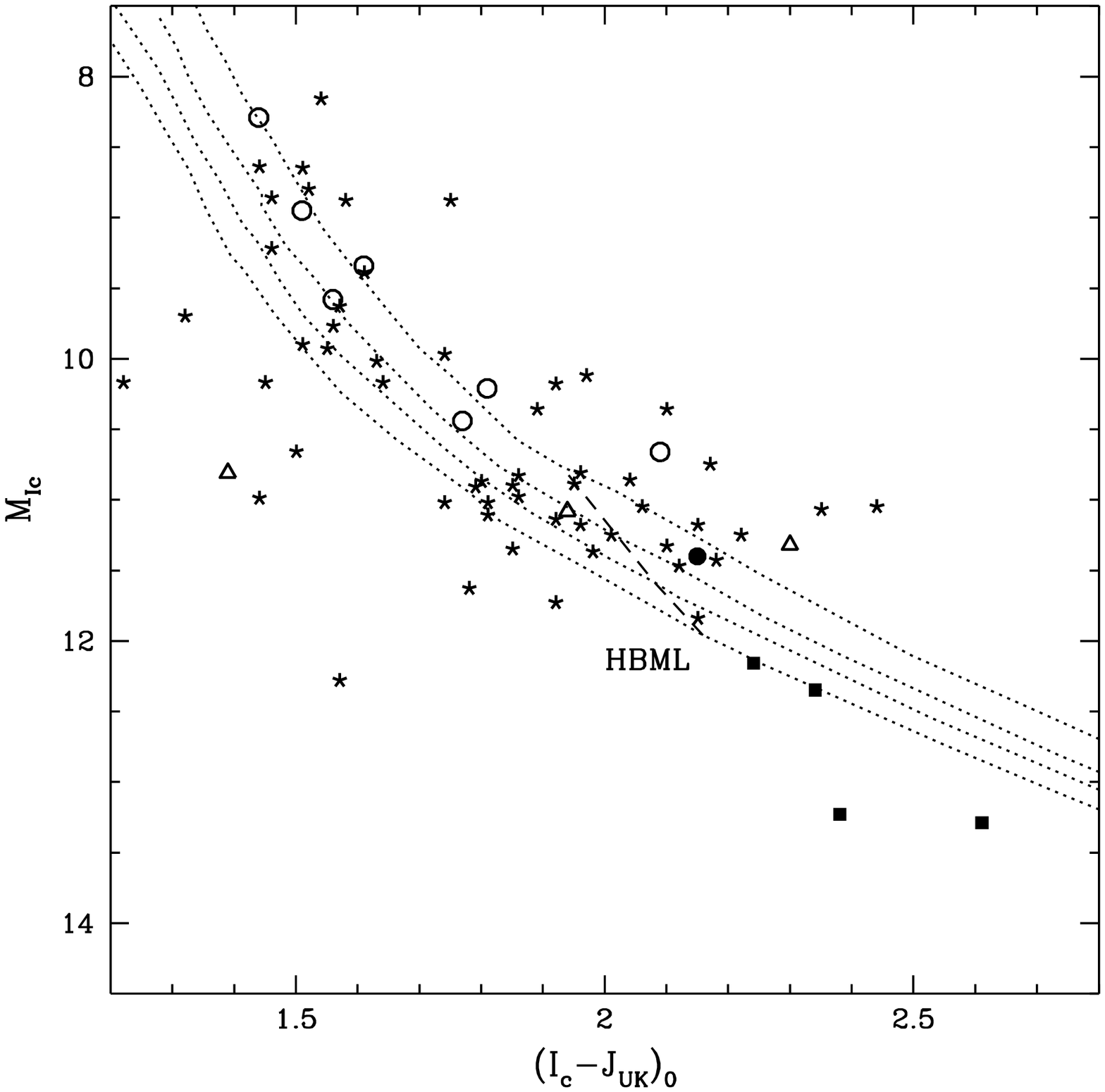}

\figcaption[magcol.eps]{\label{fig4} The absolute $I_c$ magnitude vs 
$(I_c-J)_0$ color sequence in $\alpha$~Persei (open symbols) and the Pleiades  
(five point stars). The open triangles are the  $\alpha$~Persei radial 
velocity nonmembers. The filled squares are Pleiades brown dwarfs with 
lithium detections and filled circle is AP270. The dotted lines are theoretical isochrones for 30, 50, 70 and 100~Myr. The predicted location of the Hydrogen Burning Mass Limit (0.075~M$_\odot$) is marked with a dashed line.}

\plotone{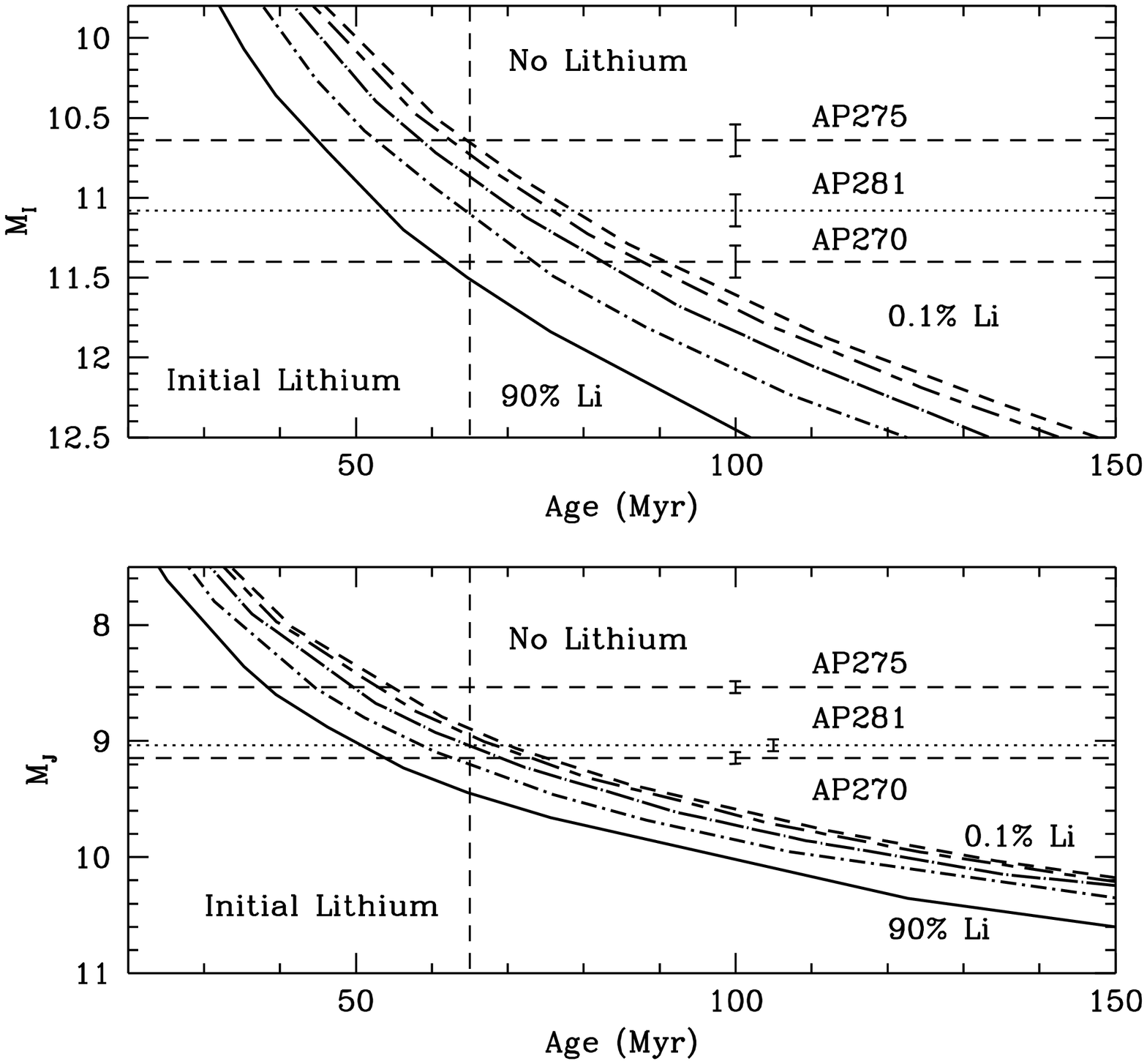}

\figcaption[jliap.eps]{\label{fig5}  
The predicted time evolution of lithium depletion as a function of absolute magnitude in the $I_c$ (upper panel) and $J$ (lower panel) filters. The solid lines are the locus of objects that have preserved 90\% of their initial lithium abundance, the dotted-short dashed line represent 50\% preservation, the dotted-long dashed lines are for 10\%, long dashed-short dashed for 1\% and short dashed for 0.1\%. The dotted lines show the absolute magnitudes 
of AP275 and AP270. The vertical dashed line is at our adopted age for the cluster of 65~Myr.}

\plotone{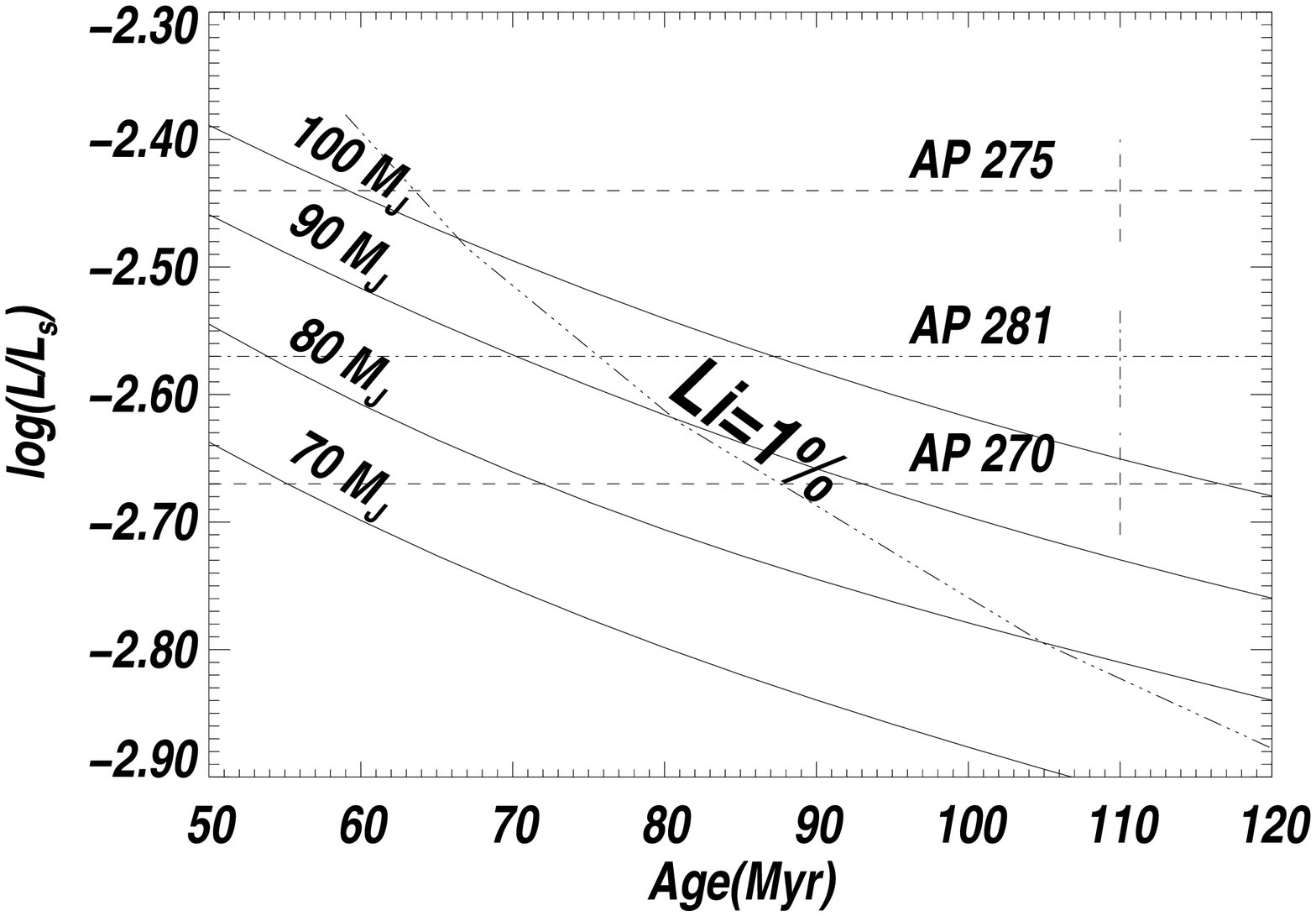}

\figcaption[lilum.eps]{\label{fig6}  
The predicted time evolution of lithium depletion as a function of absolute bolometric magnitude. The solid lines are the cooling curves for objects of the indicated masses. The dash-dot line is the expected luminosity as a function of age where lithium should be spectroscopically depleted. The horizontal dashed lines are our adopted luminosities for the members AP270 and AP275 (with errors from the uncertain distance modulus indicated). The horizontal dash-dot line is for the possible member AP281. }

\end{document}